\documentclass{emulateapj}

\usepackage{amsmath}
\usepackage{amssymb}
\usepackage{graphicx}
\usepackage{color}
\usepackage{natbib}
\usepackage{epsfig}
\usepackage{ulem}

\newcommand{\lSect}[1]{{\label{sec:#1}}}
\newcommand{\lFig}[1]{{\label{fig:#1}}}
\newcommand{\lEq}[1]{{\label{eq:#1}}}

\def\gtaprx {\lower .1ex\hbox{\rlap{\raise .6ex\hbox{\hskip .3ex
	{\ifmmode{\scriptscriptstyle >}\else
		{$\scriptscriptstyle >$}\fi}}}
	\kern -.4ex{\ifmmode{\scriptscriptstyle \sim}\else
		{$\scriptscriptstyle\sim$}\fi}}}
\def\ltaprx {\lower .1ex\hbox{\rlap{\raise .6ex\hbox{\hskip .3ex
	{\ifmmode{\scriptscriptstyle <}\else
		{$\scriptscriptstyle <$}\fi}}}
	\kern -.4ex{\ifmmode{\scriptscriptstyle \sim}\else
		{$\scriptscriptstyle\sim$}\fi}}}

\newcommand{\cutt}[1]{\textcolor{blue}{}}

\newcommand{\Ms}{{\ensuremath{\mathrm{M}_{\odot} }}}
\newcommand{\yr}{{\ensuremath{\mathrm{yr}}}}
\newcommand{\Msy}{{\ensuremath{\Ms\,\yr^{-1}}}}
\newcommand{\RS}{{\ensuremath{R_{\mathrm{S}}}}}
\newcommand{\rS}{{\ensuremath{r_{\mathrm{S}}}}}
\newcommand{\K}{{\ensuremath{\mathrm{K}}}}
\newcommand{\gcc}{{\ensuremath{\mathrm{g}\,\mathrm{cm}^{-3}}}}

\newcommand{\FIGFF}[2]{{\ref{fig:#2}{#1}}}

\newcommand{\FIG}[2]{{Fig.~\FIGFF{#1}{#2}}}
\newcommand{\Fig}[1]{{\FIG{}{#1}}}

\newcommand{\Eqref}[1]{{\ref{eq:#1}}}
\newcommand{\Eqff}[1]{{\Eqref{#1}}}

\newcommand{\Eq}[1]{{Eq.~\Eqff{#1}}}

\newcommand{\Leg}[1]{{\textit{#1}}}

\newcommand{\kepler}[1]{{{\sc Kepler}}}

\begin{document}

\title{On the Maximum Mass of Accreting Primordial Supermassive Stars}

\author{
  T.\ E.\ Woods\altaffilmark{1},
  Alexander Heger\altaffilmark{1,2,3},
  Daniel J.\ Whalen\altaffilmark{4},
  Lionel   Haemmerl\'{e}\altaffilmark{5}, and
  Ralf S.\ Klessen\altaffilmark{5}}

\altaffiltext{1}{Monash Centre for Astrophysics, School of Physics and Astronomy, Monash
University, VIC 3800, Australia}

\altaffiltext{2}{University of Minnesota, School of Physics and Astronomy, Minneapolis, MN 55455,
USA}

\altaffiltext{3}{Shanghai Jiao-Tong University, Department of Physics and Astronomy, Shanghai
200240, P.~R.~China}

\altaffiltext{4}{Institute of Cosmology and Gravitation, University of Portsmouth, Dennis Sciama
Building, Portsmouth PO1 3FX, UK}

\altaffiltext{5}{Universit\"at Heidelberg, Zentrum f\"ur Astronomie, Institut f\"ur Theoretische
Astrophysik, Albert-Ueberle-Str. 2, 69120 Heidelberg, Germany}

\begin{abstract}

Supermassive primordial stars are suspected to be the progenitors
of the most massive quasars at $z\sim6$.  Previous studies of such
stars were either unable to resolve hydrodynamical timescales or
considered stars in isolation, not in the extreme accretion flows in
which they actually form.  Therefore, they could not self-consistently
predict their final masses at collapse, or those of the resulting
supermassive black hole seeds, but rather invoked comparison to simple
polytropic models.
Here, we systematically examine the birth, evolution and collapse of
accreting non-rotating supermassive stars under accretion rates of $0.01-10\,\Msy$
using the stellar evolution code \kepler{}.  Our approach includes
post-Newtonian corrections to the stellar structure and an adaptive
nuclear network, and can transition to following the
hydrodynamic evolution of supermassive stars after they encounter the
general relativistic instability.
We find that this instability triggers the collapse of the star at masses of
$150,000-330,000\,\Ms$ for accretion rates of $0.1-10\,\Msy$,
and that the final mass of the star scales roughly logarithmically with the
rate. The structure of the star, and thus its stability against collapse, is sensitive to the treatment of convection, and the
heat content of the outer accreted envelope. Comparison with other codes suggests differences here may lead to small deviations in the evolutionary state of the star as a function of time, that worsen with accretion rate. Since the general relativistic instability leads to the immediate death of these stars, our models place an upper limit on the masses of the first quasars at birth.
\end{abstract}


\maketitle

\section{Introduction}

The possible existence of ``supermassive'' stars, with
$M\gtrsim10^4\,\Ms$, has been suggested since the early 1960s
\cite[e.g.,][]{iben63,fowler64}.
Only recently, however, have they been suspected to be necessary to
explain the formation of at least the most massive quasars found at
$z\gtrsim6$
\citep[with $M\sim10^9\,\Ms$, e.g.,][]{mort11,wu15}.  In particular,
lower-mass black holes formed from more typical Population III (Pop
III) stars could not have sustained the high accretion rates needed to grow to
such masses by this time (\citealt{wan04,pm11,wf12}; though
hyper-Eddington accretion rates may be possible, see, e.g.,
\citealt{pez16}).

In the supermassive star scenario, a primordial halo grows to masses
of $10^7-10^8\,\Ms$ without ever having formed a star, most likely
because it is exposed to a strong Lyman-Werner UV field from nearby
star forming regions \citep{agarw12,dfm14}. This destroys $\rm{H}_{2}$ and 
prevents the early collapse and fragmentation of the cloud into lower mass objects.  
In the absence of cooling by molecular hydrogen lines, the gas reaches 
temperatures of $\approx 8000$K, at which point cooling by line-emission
from collisionally excited atomic hydrogen becomes activated.
As the cloud collapses isothermally, the accretion rate scales with
the cube of the sound speed, with atomically-cooled halos permitting
catastrophic infall rates of $0.01-10\,\Msy$
\citep{wta08,rh09,sbh10,whb11,choi13,latif13a}. Other scenarios have
also been proposed which could lead to such halos
\citep[e.g.,][Yoshida et al., submitted]{io12,ivk15}. A candidate
direct collapse black hole (DCBH) has now been discovered: CR7, a
Ly-$\alpha$ emitter at $z=6.6$ \citep{cr7,til15b,agarw15b}.

In the simplest case, this produces a $10^4-10^5\,\Ms$ star at the
center of the halo, which will collapse directly to a black hole via
the general relativistic (GR) instability \citep{iben63,chandra64}.
This arises because general relativity requires a slightly greater
adiabatic exponent than that of the radiation pressure dominated gas
with $\Gamma_1=4/3$ in order for pressure support to stabilize the star
against radial pulsations.  For an $n=3$ polytrope, this occurs when

\begin{equation}
\Gamma_1-\frac43\lesssim1.12\,\frac\RS{R}\lEq{criterion}\;,
\end{equation}

\noindent where $\RS=2\,GMc^{-2}$ is the Schwarzschild radius
of the stars, and $\Gamma_1=\left(\partial\ln
P/\partial\ln\rho\right)_\mathrm{ad}$ is the adiabatic exponent
\citep{fowler64}.  For radiation-pressure dominated stars ($\beta =
P_\mathrm{gas}/P_\mathrm{tot} \ll 1$), we have the approximate
relation:
\begin{equation}
\Gamma_1 \approx \frac43 + \frac\beta6
\end{equation}
\noindent where $\beta\approx4\,k_\mathrm{B}/\left(\mu\,s\right)$,
$\mu$ is the mean molecular weight, and $s/k_\mathrm{B}$ is the
entropy per baryon.

\cite{fuller86} were the first to simulate the full evolution of a
supermassive star through the hydrodynamic collapse due to the GR
instability, but considered only the case of monolithically-formed
stars, that is, beginning their calculations with initially
supermassive models.  More recently, three studies have examined the
more realistic case, following the growth of a supermassive Pop III
protostellar core given the accretion rates expected in line-cooled
halos.  \citet{hos13} followed the growth of supermassive Pop III stars
up to $\sim10^5\,\Ms$ at several constant accretion rates and found
that they remain red and cool until they reach a few $10^4\,\Ms$.
\citet{sak15} studied the evolution of such stars in clumpy accretion
scenarios and found that the protostar could become intermittently
blue and hot at low masses but eventually evolved onto a redder,
cooler track.  Finally, \citet{ume16} included post-Newtonian
corrections in their stellar evolution calculations,
considering a sparse sample of accretion rates.  All three studies
employed stellar structure codes that were not hydrodynamical and
therefore could not follow the collapse of such stars by the GR
instability.  Thus, they could not independently determine the final
mass of the star, in order to estimate the supermassive black hole
(SMBH) seed mass at birth, but rather compared the structure of their
models with the above criterion, which is strictly valid only for
polytropes.

We have now modeled the birth, evolution and collapse of supermassive
Pop III stars to DCBHs with the one-dimensional (1D) implicit
hydrodynamics and stellar evolution code \kepler{}, which includes
post-Newtonian corrections to the structure of the star allowing it to
capture the GR instability.  \kepler{} can transition to resolving
hydrodynamic timescales should the stellar model under consideration
encounter a dynamical instability.  This allows us to model the
complete evolution of the star until the end of its life, and through
the onset of its collapse to a black hole.  Our approach includes
accelerated nuclear burning driven by collapse that might be capable
of slowing or reversing it.  We obtain final masses for these stars
over the range of central collapse rates in atomically cooled halos
found in high resolution cosmological simulations.  Our \kepler{}
models are described in Section~2 and the evolution and final masses
of Pop III SMS are discussed in Section~3. We consider the
implications of our simulations for the properties of high redshift
quasars in Section~4.

\section{Numerical Method}

\subsection{\kepler{}}

\kepler{} \citep{Weaver1978,fuller86,Woosley2002} is a one-dimensional
(1D) Lagrangian hydrodynamics and stellar evolution code with nuclear
burning and mixing due to convection.  \kepler{} can use a 19-isotope
``APPROX'' nuclear reaction network \citep{Weaver1978} which is
adequate for many stellar evolution situations. However, for the
studies presented here, we use an adaptive full nuclear reaction
network implicitly coupled to the hydrodynamics \citep{WH04}.
\kepler{} uses an equation of state similar to, and compatible with,
the Helmholtz equation of state \citep[EOS;][]{ts00}, which includes
contributions from degenerate and non-degenerate relativistic and
non-relativistic electrons, electron-positron pair production, and
radiation.

The star is partitioned into a maximum of 1982 zones in mass.
Consistent with the general consensus that massive Pop III stars do
not lose much mass over their lives, we turn off mass loss in our
models, assuming that any mass loss will be negligible relative to
the rapid accretion rates considered here.  In particular, pulsational
mass loss is also unlikely to be significant for \textit{accreting}
supermassive stars \citep{hos13}.  In order to capture the physics of
the GR instability, we use the 1st order post-Newtonian approximation to the Tolman-Oppenheimer-Volkoff correction
to the structure of the star \citep{zeld71,kip12} as described in
Section~2.1 of \citet{chen14b}.

The accreted material is assumed to be of the same primordial
composition as the cloud from which the original seed of the star has
formed.  We assume cold accretion \citep[following, e.g.,][]{hos13},
which means material is accreted with the surface entropy. In order to
accurately follow the transfer of heat through the accreted envelope,
we switch from a wholly Lagrangian to a semi-Eulerian treatment in the
outermost part of the star.  We make the transition at a specified total optical
depth measured inward from the surface, here chosen to be $10^6$ in
order to ensure the outer surface layers remain well-resolved.  In
practice, this amounts to adding a homologous term to the
gravitational energy release in the outer layers of the star, in order
to account for the work done by compression of the accreted material.
In principle, some fraction of the accretion luminosity should be 
released at the accretion shock, a part of which should contribute to 
an additional heating term at the surface of the star. This was found 
to have a negligible effect on the evolution of supermassive stars, assuming
any fraction $0 \leq \eta \leq 1$ of this luminosity is added to the bottom 
of the photosphere \citep{hos13}. A more detailed treatment may change this
picture slightly, however given the current uncertainties in the formation
of supermassive stars, we ignore this term in the present work.

\subsection{Initial Conditions}
\lSect{initial}

Due to the difficulty in achieving numerically stable results when
beginning at lower stellar masses, we initialize all of our models as
$10\,\Ms$, chemically homogeneous, $n=3$ polytropes.  The initial
central density, $\rho_\mathrm{c}$, is set to $10^{-3}\,\gcc$, giving a
central temperature $T_\mathrm{c} = 1.2\times 10^6\,\K$ capable of
sustaining deuterium-burning.
We assume a baryon-to-photon ratio, $\eta$, equal to $6\times10^{-10}$
in the primordial gas \citep{cyb01,cyb02} which is consistent with the results of Big Bang Nucleosynthesis, and is consistent with initial mass fractions of hydrogen, $^4$He, $^3$He, $^{2}$H, and Li of $\approx0.75$, $\approx0.25$, $\approx2.1\times 10^{-5}$, $\approx 4.3\times 10^{-5}$, and $\approx 1.9\times 10^{-9}$ respectively (B.\ D.\ Fields, priv.\ comm.).

\begin{figure*}[t]
\begin{center}
\begin{tabular}{cc}
\epsfig{file=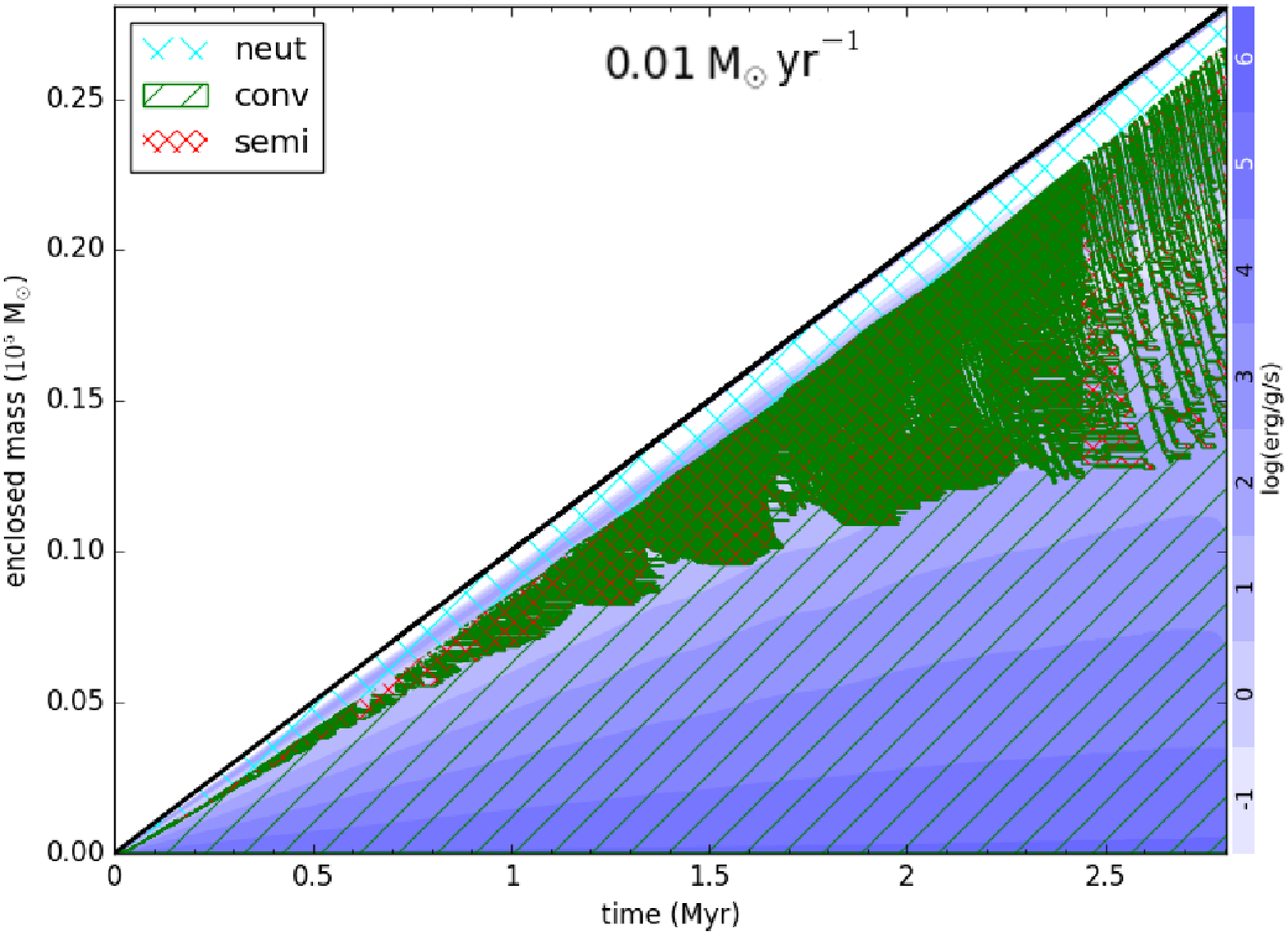,width=0.5\textwidth} &
\epsfig{file=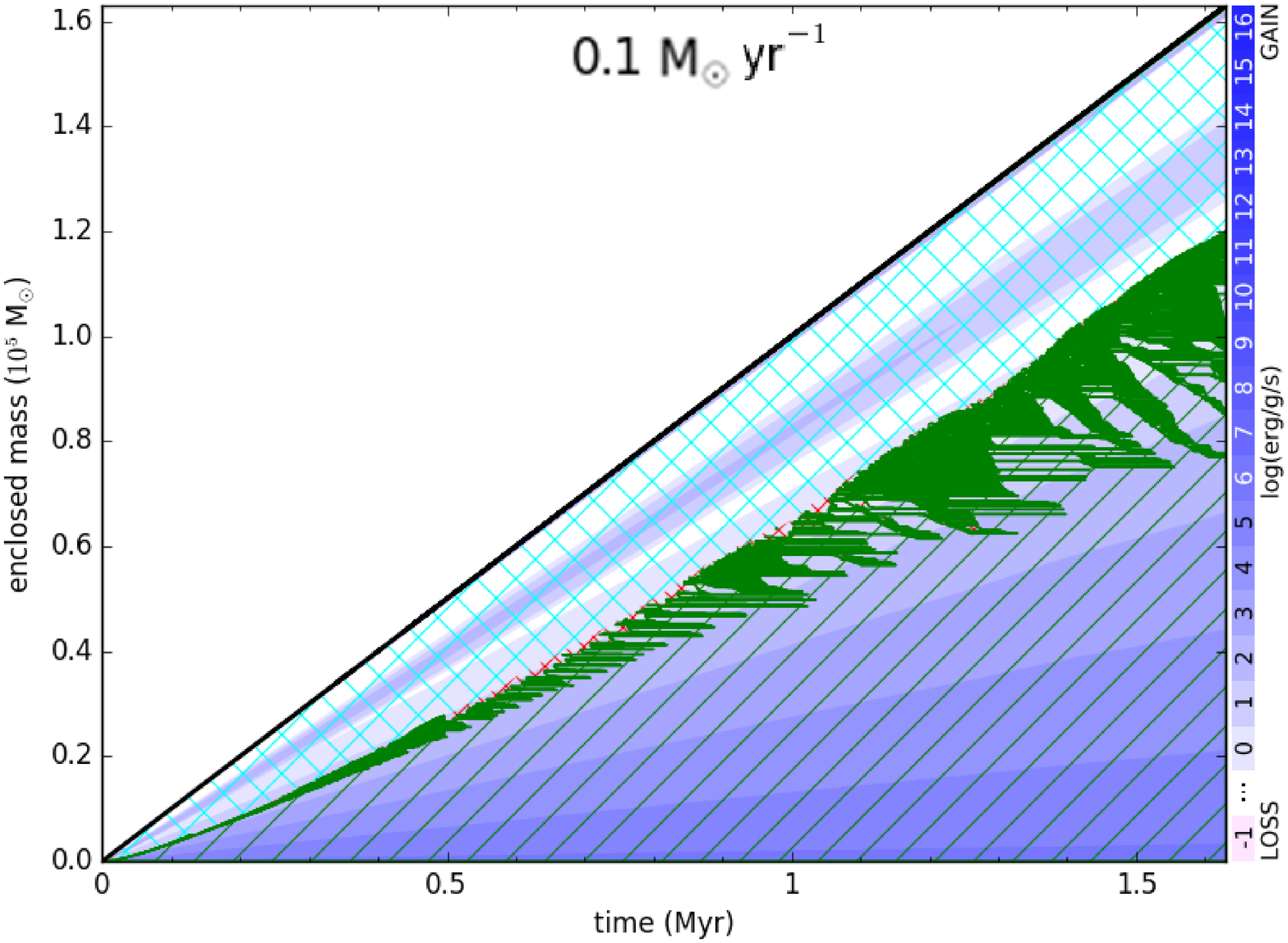,width=0.5\textwidth} \\
\epsfig{file=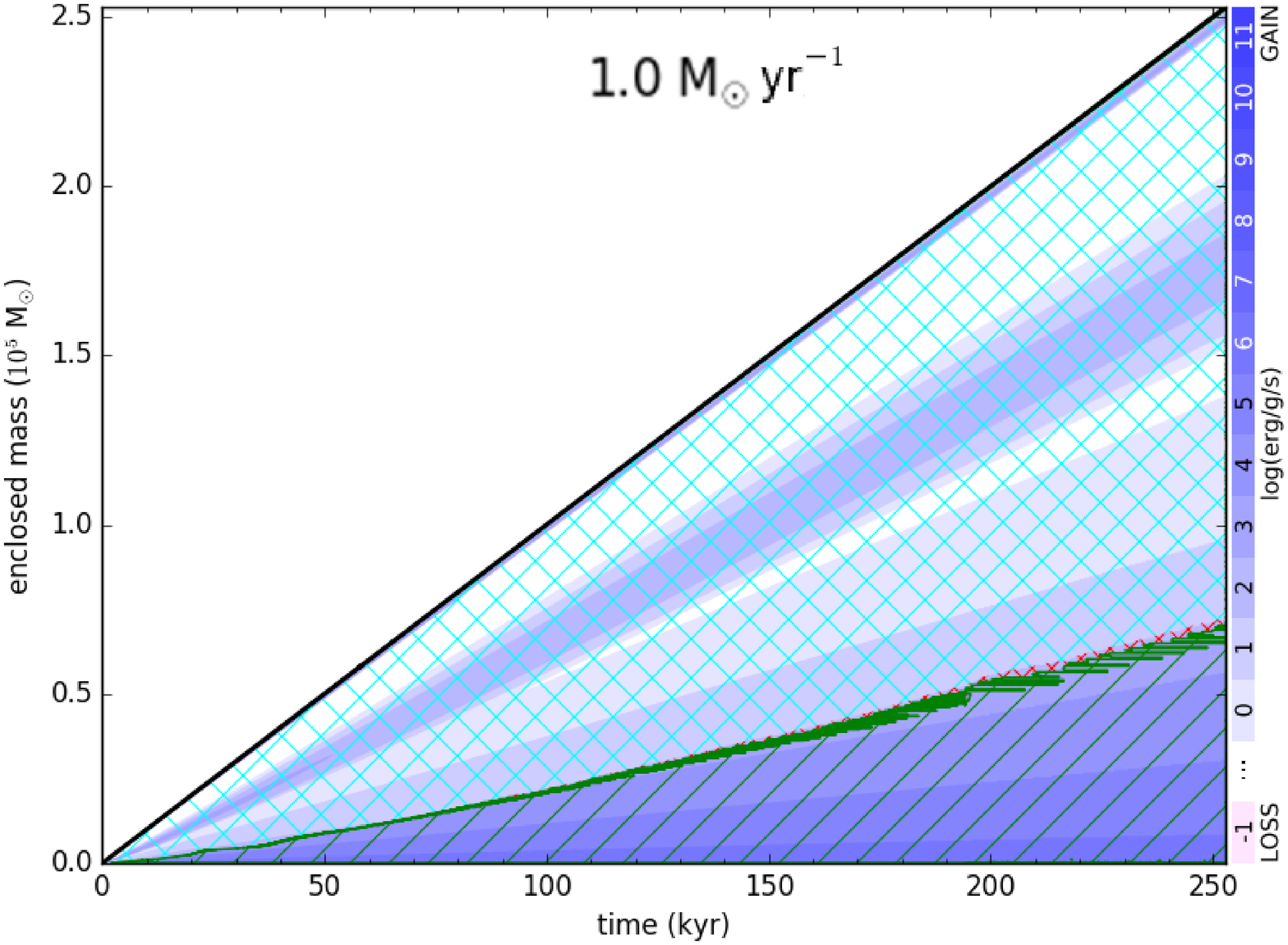,width=0.5\textwidth} &
\epsfig{file=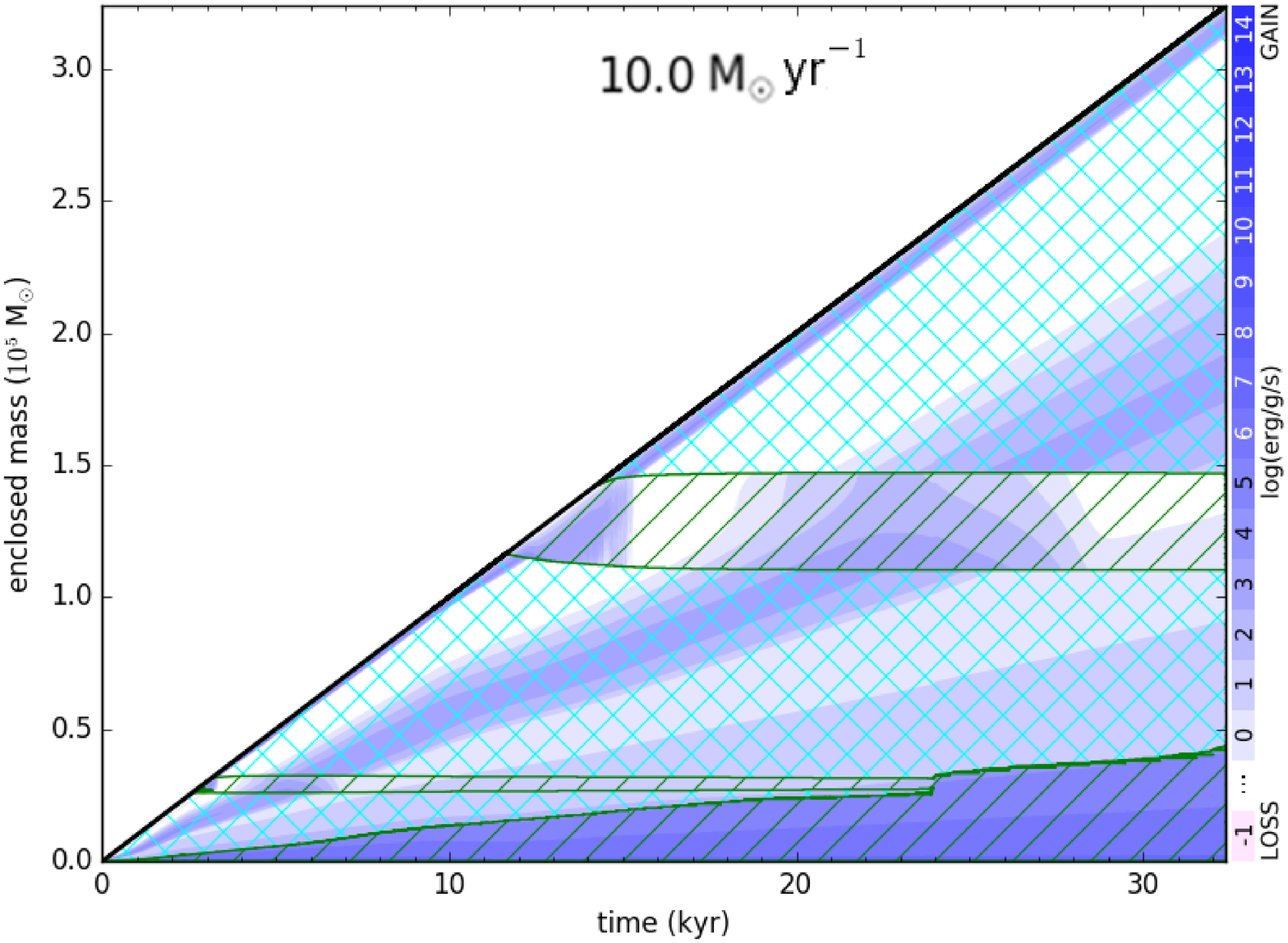,width=0.5\textwidth}\\
\end{tabular}
\end{center}
\caption{Kippenhahn diagrams of the interior of the star accreting at $0.01\,\Msy$ (\Leg{top left})
  $0.1\,\Msy$ (\Leg{top right}), $1\,\Msy$ (\Leg{bottom left}), and $10\,\Msy$
  (\Leg{bottom right}).  Hashing denotes convective regions (\Leg{green}, ``conv'')
  and neutral (\Leg{turquoise}, ``neut'') regions, with 
  \Leg{solid green lines} denoting
  the boundaries of convective regions.  Regions without hashing are
  radiative. Specific energy generation rate at each point in the
  star is indicated by the color axis. Note that we have truncated
  the plot for the $0.01\,\Msy$ case at approximately the end of H-burning,
  for clear comparison with the stellar structure of the other models
  at the relevant evolutionary stage.}  \lFig{kipp}
\end{figure*}

We consider seven constant accretion rates uniformly spaced in
$\log\dot{M}$ from $0.01\,\Msy$ to $10\,\Msy$ that span the infall
rates expected in Lyman-cooled halos based on cosmological
simulations.
The lowest, $0.01\,\Msy$, corresponds to the maximum accretion rate
due to H$_2$ cooling in less massive halos.

\section{The Evolution of Accreting Supermassive Stars}

The evolution of Pop III SMSs depends sensitively on the balance
between the accretion of new material, the nuclear evolution of the
core, and the thermal relaxation of the star.  In \Fig{kipp}, we show
the evolution of the stellar structure and energy generation with time
in four Kippenhahn diagrams for the $0.01\,\Msy$, $0.1\,\Msy$, $1\,\Msy$, and
$10\,\Msy$ cases.  Each model develops a distinct inner convective
core and outer, high-entropy envelope, as expected
\citep[e.g.,][]{begel10}.  There is some uncertainty in the precise
outer boundary of the convective core, becoming most pronounced for
the lowest accretion rates, where the difference in entropy between
core and envelope is lowest. This in turn is a consequence of the greater thermal
relaxation of the envelope at lower accretion rates. This ambiguity is an inescapable
consequence of the simplified 1D treatment of convection utilized in 1D stellar evolution
codes.  At relatively low accretion rates ($\lesssim 10^{-1.5}\,\Msy$), the fate of the star is
principally set by the hydrogen nuclear-burning lifetime of the star.
At higher accretion rates, the star encounters the
GR instability while still on the hydrogen-burning main sequence, due
to the growth of the convective core.

As a representative example, we consider a star that grows at a rate
of $1\,\Msy$.  It reaches a central temperature of $10^8\,\K$
after $\sim 1,485$ yr at a density of $20\,\gcc$.  This allows the
core to produce a rapid spike in CNO elements through the
triple-$\alpha$ process, catalyzing hydrogen-burning until the core is
stabilized against further contraction (see \Fig{central}).  The star
develops a small convective core shortly after this
($\approx 2,000\,\yr$), with an extended, high-entropy envelope.  The
core grows approximately linearly with time until encountering the GR
instability when the total mass of the star reaches $\sim 300,000\,\Ms$
(that is, after $\sim 300,000\,\yr$).  The inner convective core at
this time has reached $\approx 60,000\,\Ms$ and a helium fraction of
$\sim\,50\,\%$.  After the onset of collapse, infall velocities
quickly reach a few percent of the speed of light.  We halt our
calculation here, before the collapse becomes strongly relativistic,
as our models include only post-Newtonian corrections to gravity but no
relativistic hydrodynamics.

\begin{figure}
\begin{center}
\epsfig{file=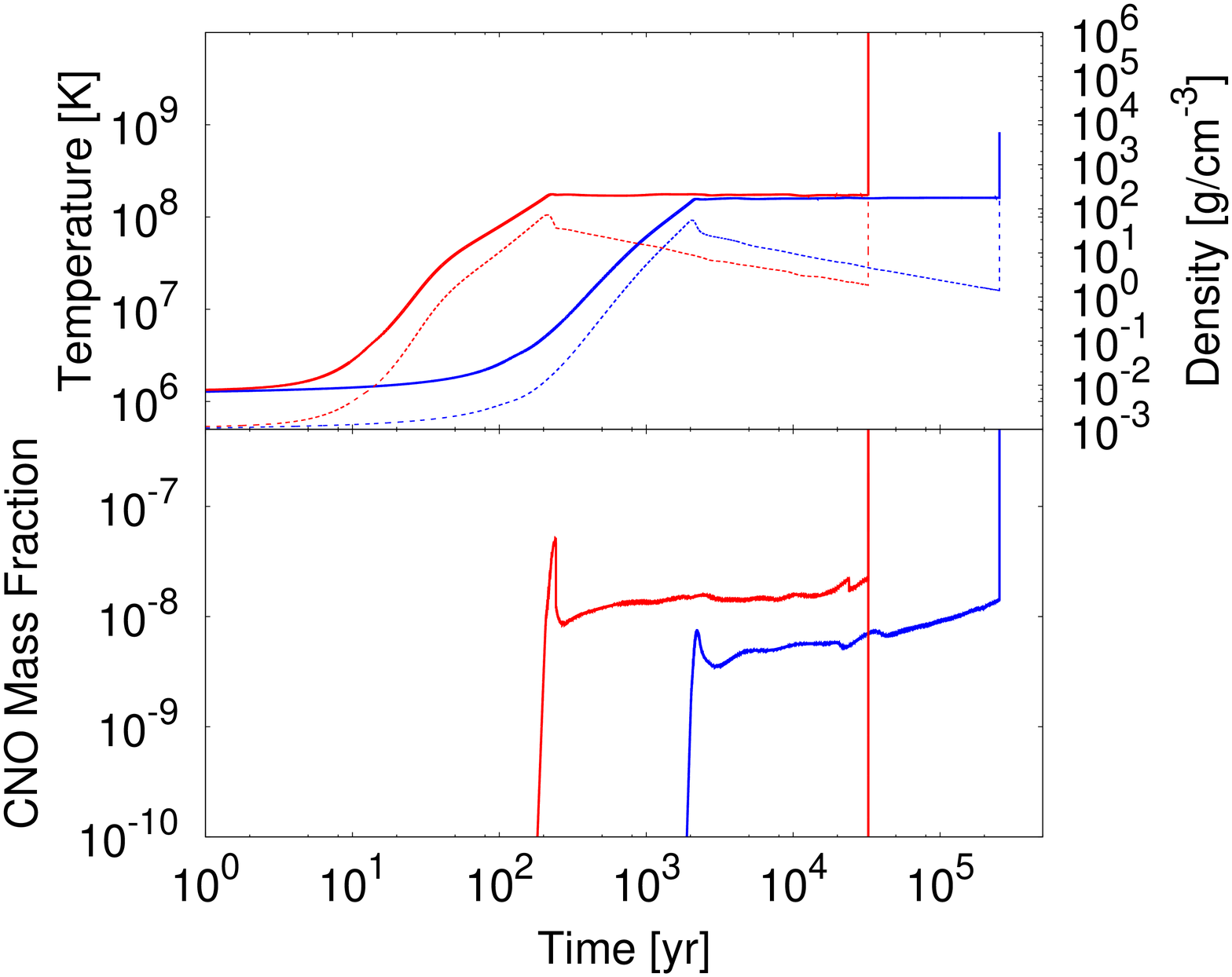,width=\columnwidth}
\end{center}
\caption{Central densities (\Leg{upper panel, dotted lines}),
  temperatures ({\it upper panel, solid lines}), and CNO abundances
  ({\it lower panel}) for stars accreting at $1\,\Msy$ (\Leg{blue}),
  and $10\,\Msy$ (\Leg{red}).}\lFig{central}
\end{figure}

\begin{figure}
\begin{center}
\epsfig{file=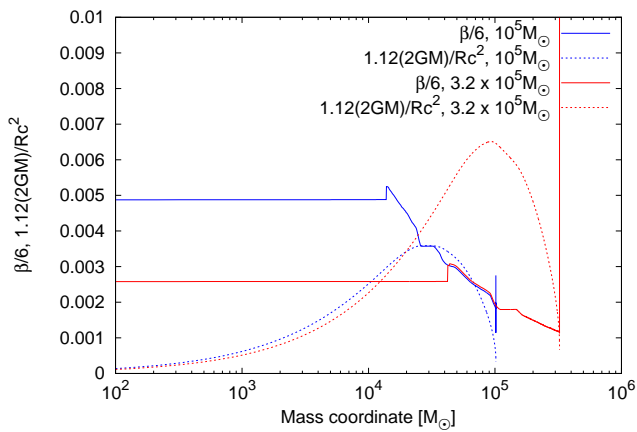,width=\columnwidth}
\end{center}
\caption{Comparison of $\Gamma_1-4/3\approx\beta/6$ (\Leg{solid
    lines}) with the $n=3$ polytropic criterion for instability
  (\Leg{dashed lines}) for our $10\,\Msy$ model at $\approx 10^5\,\Ms$
  (\Leg{blue lines}) and $\approx3.2\times10^5\,\Ms$ (\Leg{red
    lines}).  The latter mass is reached shortly before
  collapse. }\lFig{crit}
\end{figure}

\begin{figure*}
\begin{center}
\epsfig{file=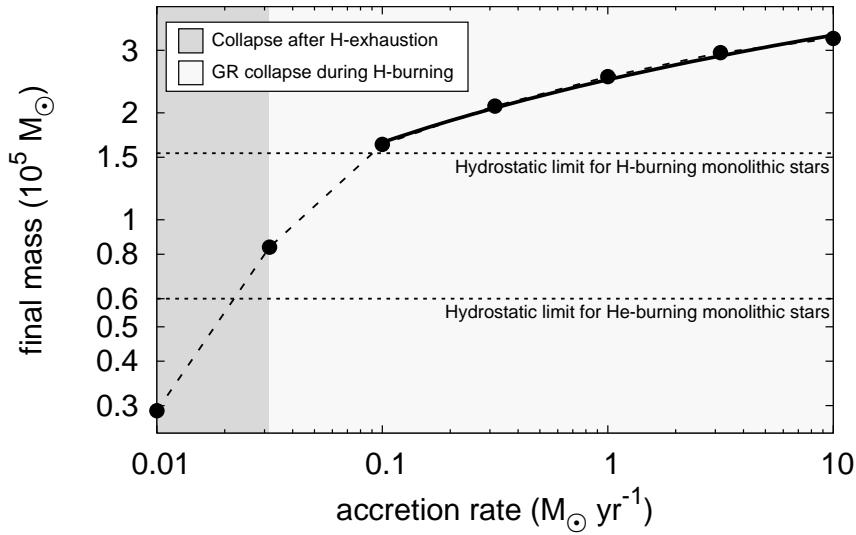,width=0.65\linewidth,clip=}
\end{center}
\caption{Final masses as a function of accretion rate.  \Leg{Black
    circles} denote individual models undergoing constant accretion,
  with the trend given by the \Leg{black dashed line}.  The \Leg{solid
    black line} plots the fit formula \Eq{fit}.  Also shown are the
  limiting masses for hydrostatic hydrogen and core helium burning for
  monolithically-formed SMSs.
}
\lFig{mdotmfinal}
\lFig{halo}
\vskip0.5in
\end{figure*}

For all of our models which accrete above $\gtrsim10^{-1.5}\,\Msy$,
the evolution follows similarly to that described above, except that
their longer lifetimes prior to collapse via the GR instability allow
them to reach core helium fractions of up to $99\,\%$.  For accretion
rates of $0.01\,\Msy$ and $10^{-1.5}\,\Msy$, we find that our stellar
models reach total masses of only $32,000\,\Ms$ and $84,000\,\Ms$,
respectively, before exhausting hydrogen in their cores.  Our
$10^{-1.5}\,\Msy$ model collapses immediately upon core
hydrogen-exhaustion, whereas our $0.01\,\Msy$ model survives to
silicon burning before the collapse of the core.

Turning to the highest accretion rates, the most strikingly different
feature is the emergence of additional convective zones in the
envelope, driven by instabilities arising near the surface and accompanied
by radial pulsations.  These convective zones are also seen in
\cite{ume16}, as is evident from the small flat plateaus in the
entropy profiles in the envelope of their $10\,\Msy$ model, as seen in
their Fig.~3 (top left). Therefore this cannot be a source of any
discrepancy between our results and theirs regarding the evolution and fate of the star. Our model encounters the GR instability at a
convective core mass $\approx42,000\,\Ms$, similar to the point at
which they cease their calculations after entering the pair-unstable
regime.  This, however, occurs at a total mass of only
$\approx330,000\,\Ms$ in our models, contrasted with their
$800,000\,\Ms$. Indeed, evidently the $10\,\Msy$ model of \cite{ume16},
is far less evolved than our $10\,\Msy$ model at any fixed total mass.
The primary reason for this is unclear, but must arise through some
difference in the entropy profile and convective stability of the star.
This emphasizes the uncertainty in the treatment of
convection and the cooling of the outer envelope. \cite{hae17} use a
third stellar evolution code and find final masses similar to ours at
low accretion rates \citep[as do][]{ume16}, but intermediate between
those presented here and that of \cite{ume16} for $10\,\Msy$.

In \Fig{crit}, we visualize the criterion for the onset of the GR
instability as derived for an $n=3$ polytrope \citep[\Eq{criterion},
  see][]{chandra64}, plotting $\beta/6$ and $1.12\,\rS/r$ against mass
coordinate, $m = m\left(r\right)$.  Here
$\rS=\rS\left(m\right)=2\,Gmc^{-2}$ is the ``local Schwarzschild
radius'' of the enclosed mass $m$.  As found by \cite{ume16}, long
before the onset of collapse $\beta/6>1.12\,\rS/r$ throughout the
convective core and indeed, almost the entirety of the envelope,
whereas near the point of collapse the region where this condition is
violated has passed into the isentropic core.  The moment when this is
satisfied at the surface of the core, however, does not mark the onset
of collapse in our models, nor did it mark the endpoint in the
calculations of \cite{ume16}, as in the final panel of their Fig.~3
the intersection of the two curves is already well within the core
boundary.  This is reasonable, as the entirety of the star is clearly
not well-described by a $n=3$ polytrope (given the accreted envelope),
nor is the core alone reasonably approximated as such, although $\beta
\ll 1$. We then conclude that, even when applied
only to the isentropic core, the \cite{chandra64}
criterion does not adequately describe the mass needed to trigger collapse in
realistic supermassive stars. A modified polytropic approximation \citep{begel10}
 may be able to better capture the critical pressure gradient needed, but this should
also depend on the accretion and evolutionary history of the star, rendering it
less useful.
Following the hydrodynamic response of the star
is necessary then in order to be certain of the moment the GR
instability sets in.

Our results are summarized in \Fig{mdotmfinal}, where the final
masses and evolutionary characteristics of supermassive stars are
given as a function of the accretion rate.  For accretion rates
$\lesssim10^{-1.5}\,\Msy$, stars collapse only after the onset of
He-burning, while for greater accretion rates, GR collapse occurs
during core hydrogen-burning.  For accretion rates $\gtrsim0.1\,\Msy$,
the final mass at collapse varies as:

\begin{equation}
M_\mathrm{SMS,final}\approx\left[0.83\,\log_{10}\left(\frac{\dot{M}}{\Msy}\right)+2.48\right]\times10^5\,\Ms
\lEq{fit}
\end{equation}

\noindent fitting our results to within $\approx3\,\%$.  Also shown
for reference in \Fig{mdotmfinal} are the maximum masses for
hydrostatic hydrogen- and helium-burning for monolithically-formed (as
opposed to accreting) supermassive stars, found from trial
computations beginning with supermassive, $n=3$ polytropes.

\section{Discussion and Conclusion}

We find that, in general,
supermassive stars survive well past the point where the isentropic core
satisfies the \cite{chandra64} criterion, before finally encountering the hydrodynamic
GR instability. This is a natural consequence of the much different structure of
an accreting supermassive star compared with a simple $n=3$ polytrope. Our models predict that the general relativistic instability imposes a
characteristic final total mass of $\lesssim300,000\,\Ms$ on
accreting, primordial supermassive stars, as for accretion rates above
$\approx0.1\,\Msy$ the final mass at collapse grows only
logarithmically with the accretion rate.  Such infall rates are
thought to be typical of atomically-cooled primordial composition
clouds, providing the seeds of SMBHs formed via direct collapse
including high-redshift, massive quasars.
Since BHs built up in runaway collisions are generally an order of
magnitude lower in mass \citep{dv09}, these results are likely an
upper limit on the masses of SMBHs which may have formed via direct
collapse.

This picture could still change if stellar rotation is taken into
account \citep{shibata16}  Rotation rates of up to $50\,\%$ of the critical, or
breakup, velocity at birth have been found for much lower mass Pop III
stars in numerical simulations \citep[][Haemmerle et al., in prep.]{stacy11b,stacy13}.  Rotational
mixing could alter the structure of the star and support it
against collapse up to somewhat higher masses.  This potentially has observational implications as the collapse of such stars could be
accompanied by strong gravitational wave emission
\citep[e.g.,][]{FWH01}. Rotation may or may not also lead to significant
mass loss driven by outflows during the final collapse of the star, depending
on the internal redistribution of angular momentum during collapse
\citep{FR17,Uchida17}.

The collapse of massive, atomically-cooled primordial halos allows for
extreme accretion rates sustained for a remarkable duration, far
surpassing any star formation environment encountered later in the
evolution of the Universe.  The resulting supermassive objects must
then have been among the most massive stars to have ever existed, and are
strong candidates for the progenitors of some of the first and most luminous
quasars.

\acknowledgments

AH was supported by an Australian Research Council (ARC) Future
Fellowship (FT120100363) and NSF grant PHY-1430152 (JINA-CEE).  
D. J. W. was supported by STFC New Applicant Grant ST/P000509/1.
LH,DJW, and RSK were supported by the European Research Council
under the European Community's Seventh Framework Programme (FP7/2007 -
2013) via the ERC Advanced Grant ''STARLIGHT: Formation of the First
Stars'' (project number 339177).  Part of this work was supported by
the Swiss National Science Foundation.  Work at LANL was done under
the auspices of the National Nuclear Security Administration of the
U.S.\ Department of Energy at Los Alamos National Laboratory under
Contract No.\ DE-AC52-06NA25396.

\bibliographystyle{apj}

\end{document}